\documentclass{article}
\usepackage{amsfonts}
\usepackage{amssymb}
\usepackage{amsmath}
\usepackage{amsthm}
\usepackage[english]{babel}
\begin{document}

\title{\bf On cosmic inflation in vector field theories}

\author{{\bf Alexey Golovnev}\\
{\small {\it Saint-Petersburg State University, high energy physics department,}}\\
{\small \it Ulyanovskaya ul., d. 1; 198504 Saint-Petersburg, Petrodvoretz; Russia}\\
{\small agolovnev@yandex.ru}}
\date{}

\maketitle

\begin{abstract}

We investigate the longitudinal ghost issue in Abelian vector inflation. It turns out that, within
the class of Lorentz-invariant vector field theories with three degrees of freedom and
without any extra (scalar) fields,
the possibilities are essentially exhausted by the classical solution due to Larry Ford
with an extremely flat potential which doesn't feel the fast roll of its argument. And, moreover,
one needs to fulfill an extra condition on that potential in order to avoid severe gradient
instability. At the same time, some Lorentz-violating modifications are worth to be explored.

\end{abstract}

\vspace{1cm}

\section{Introduction}

Inflation is one of the most successful attempts to understand the reasons for why
our Universe is so flat, and large, and homogeneous. Inflation is usually driven
by a scalar field called inflaton. Not only does this picture solve the problems of
the standard Big Bang cosmology, but it actually provides us with a very nice origin
of tiny inhomogeneities seen in the cosmic microwave background radiation which
presumably gave rise to all the structures in the Universe. However, despite this
great success, the nature of the inflaton remains largely unknown. It is therefore
interesting to explore some other concievable types of inflationary models.

Probably,
the simplest alternative idea is to use the usual (massive) vector fields. However, it is easy to see that vector fields
have two major problems: generically they do not satisfy the slow-roll conditions and do
induce too large an anisotropy of expansion. In the pioneering work
by L. Ford \cite{Ford} we find two possible solutions of the slow-roll problem.
First, in a version of new inflationary scenario the effect of tachyonic vector
field mass can balance (if fine-tuned) the inflationary dilution of the vector field potential
energy (see also \cite{Armen,Dim}).
Second, the vector field potential in chaotic inflation can be
taken so extremely flat that it would be practically insensitive to the
fast roll of its argument, $A_{\mu}A^{\mu}$. The anisotropy problem can be cured by
some specific configurations of vector fields (vector triples) \cite{Hosotani, Galt, Bertolami, Armen, GMV}
or with a large number of randomly oriented independent non-interacting fields \cite{GMV}.
It doesn't actually quite work for large fields models, and strong anisotropies do generically develop
in the chaotic inflationary regimes \cite{GMV2, me}. Nevertheless, we would ignore the anisotropy
problem in this paper and concentrate on enforcing the slow-roll conditions in a reasonable
way. The standard approach \cite{GMV} (apart from explicit tachyonic mass)
is to prevent the vector fields from decaying with a
non-minimal coupling to gravity of the $RA_{\mu}A^{\mu}$ form, see also \cite{Harko,Mota}. It perfectly works at the
background level. However, the problem is that effectively this coupling acts, of course, as
nothing else but tachyonic
mass. And the tachyonic mass for vector fields means that the longitudinal modes are ghosts
(at sub-Hubble scales) and lead to violent instabilities \cite{Peloso1, Peloso2, Peloso3, me, Esposito}.

Let us summarize the known stability problems in vector inflation. Probably, the first to be
reported was the problem of anisotropic instability in chaotic-type models. Even at the background level,
it was clear \cite{GMV} that one can not start vector inflation from arbitrarily high values of
$N$ randomly oriented vector fields because the isotropic solution requires a
statistical cancellation of pretty large anisotropic terms; and for the large vector field values
the anisotropic statistical fluctuation dominates over the mean isotropic quantity. This simple
observation led to conclusion \cite{GMV} that, for the mass-term potential, one can generically have
$\sim 2\pi\sqrt{N}$ e-folds of nearly isotropic inflation. It provides a first hint that it would be quite expectable
if even very small anisotropic vector field fluctuations around the background solution are to become
(linearly) unstable. And this is actually the case. In \cite{GMV2} it was claimed that
gravitational waves generically possess a bad tachyonic instability. This analysis was incomplete as
all the couplings with other modes were unsubstantiatedly neglected (recall that the so-called decomposition
theorem is not valid in vector inflation \cite{Armen}). One could also suspect that the
exponential growth of anisotropies is just an artifact of the linear perturbation analysis in the Jordan frame.
However, in the long-wavelength limit this effect can be seen at the fully non-linear level too \cite{me}.

Another, and the most worrisome, stability problem is that of the longitudinal vector field
fluctuations. For tachyonic masses the longitudinal modes are ghosts at large momenta. We will see that,
in the class of Lorentz-invariant vector field theories, chaotic inflationary model by L. Ford is
virtually unique as a candidate for stable realization of vector inflation. But we will also see
that even among these fine-tuned ghost-free potentials, very dangerous gradient instabilities do
sometimes occur around inflationary backgrounds.

Finally, the number of degrees of freedom is ill-defined in the model with $RA_{\mu}A^{\mu}$-coupling \cite{me}.
The scalar curvature contains second space-time derivatives of the metric field, and therefore
the temporal components $A_0$ (or more precisely, one combination of $A_0$-components
of all the vector fields in the model, see \cite{me})
become dynamical, but not around the homogeneous background
for which $A_0=0$. Spatial homogeneity amounts to the strong coupling regime for the temporal
vector field modes.

Let us also note that vector inflation has been later generalized to higher
p-form fields \cite{Germani1, KMota} too. With special couplings to Ricci scalar and Ricci tensor,
these theories can also exactly mimic the usual slow-roll dynamics of a scalar inflaton. Moreover,
in \cite{Germani1} it was found that there exist duality transformations which relate 2-form models
to vector inflationary ones, and 3 forms -- to scalars. The latter allowed for a detailed analysis
of linear perturbations in the 3-form (isotropic) inflation \cite{Germani2}.
Note though that these transformations involve both curvature scalar
$R$ and $R_{\mu\nu}$ tensor in a non-trivial way, and therefore they do contain second time derivatives
of metric which leads to the higher (third) derivative terms in the energy-momentum tensor in the
dual picture. The troublesome terms do vanish in homogeneous background and correspond, of course,
to the aforementioned problem with extra ill-defined degrees of freedom. However, it is important
to understand that there is no need to mimic the scalar field evolution precisely, even
in slow-roll scenarios. We can not afford
the usual $\sim H^2$ contributions to the effective mass since they render
the Hubble friction ineffective. But any corrections of order ${\dot H}$ make no harm
in quasi-de-Sitter stage since the Hubble constant is almost constant.
Therefore, one could for example couple the vector inflaton not only to $R$ but also
to $R_{\mu\nu}$. This fact is of particular importance for three-form models which
actually admit the isotropic minimal-coupling slow-roll scenario \cite{Nunes1,Nunes2} without
the usual instability problems. At the same time, the two-form inflation shares all of the problems
typical for vector inflation including both the ghost issues \cite{Germani1} and
the catastrophic growth of anisotropies \cite{YK}.

In this paper we restrict our attention to the case of {\it Abelian} vector inflation, and in particular to
its ghost problem. Our major interest is to find out whether it is possible to overcome the
ghost problem of vector inflation without intriducing extra (scalar) fields into the model.
(Note that there is at least one model on the market which produces a stable anisotropic
inflation with vector fields coupled to a scalar (dilaton) field via the $f(\phi)F_{\mu\nu}F^{\mu\nu}$
term, see \cite{WKS1,WKS2,thor}.) Other stability problems are beyond the scope of the present work.
Therefore we consider test vector fields in a fixed background geometry.
It is enough for discussion of the vector field longitudinal mode behaviour. In this approximation,
the curvature scalar and tensor
can be treated just as (time-dependent) parameters in a minimally-coupled vector field Lagrangian.
Even apart from the time-dependence,
this effective theory may appear to be Lorentz-violating even for Lorentz-invariant models due
to possible $R_{\mu\nu}A^{\mu}A^{\nu}$ coupling. However, both the Lorentz-violating effects and
the time-dependence of effective parameters are small in quasi-de-Sitter stages. We will first discuss
only Lorentz-invariant vector field theories, and then make some comments on Lorentz-violating models.
Note also that throughout the paper we assume the spatially flat FRW-spacetime, while it is known to
be quite hard to start vector inflation in spatially curved backgrounds \cite{Chiba}.

\section{Canonical vector fields}

In vector inflation we assume that at the background level the fields are homogeneous. For
canonical massive (Proca) vector fields the equations of motion are \cite{GMV}:
$\bigtriangledown_{\mu}F^{\mu\nu}+m^2A^{\nu}=0$. Under the spatial homogeneity assumption,
the temporal component of these relation yields $A_0=0$ \cite{Ford, GMV}. And this is
generically true of many more complicated vector models too. At first glance, we have
a slow-roll regime for the spatial vector field components due to remaining equations
of motion, ${\ddot A}_i+H{\dot A}_i+m^2A_i=0$, in the FRW spacetime with metric element
$$ds^2=dt^2-a^2(t){d\overrightarrow x}^2$$
where
$a$ is the scale factor and $H\equiv\frac{\dot a}{a}$ is the Hubble constant.
However, it is not what we need because the potential
energy, $-\frac{m^2}{2}A_{\mu}A^{\mu}$, decays a bit faster than $a^{-2}(t)$. One
can introduce a somewhat more physical variable $B_i\equiv\frac{A_i}{a}$. Unfortunately, it gets
a large effective mass:
$${\ddot B}_i+3H{\dot B}_i+\left(m^2+2H^2+{\dot H}\right)B_i=0$$
which makes the Hubble friction ineffective and a slow-roll regime impossible.
It's really bad news for vector inflation, however this issue has also received much attention
due to a more mundane reason. In particular, we mean the problem of the seeds of primordial magnetic
fields in the Universe. Magnetic fields are known to exist in galaxies and in clusters of galaxies,
and at the same time it is very hard to imagine how could have any considerable magnetic seeds survived after inflation
to be used in, for example, subsequent dynamo mechanisms. We will not discuss this topics
any further, and refer the interested reader to the existing literature \cite{Turner, Rubinstein, Vittoria}.

In the pioneering paper \cite{Ford} we can find two possible realizations of the slow-roll dynamics
in vector inflation. The first idea is that one can introduce a tachyonic vector field mass in such a way that it
will almost compensate the unwanted geometrical contribution, $2H^2+{\dot H}$. A possible way to introduce
such an effective mass without too much evident fine-tuning is to invoke a particular non-minimal
coupling term, $\frac{R}{12} A_{\mu}A^{\mu}$, in the Lagrangian. (Recall that in the FRW-spacetime we
have precisely $\frac{R}{6}=-2H^2-{\dot H}$.) In any case, for a test vector field it is just a tachyonic
mass term, and one can make a St{\" u}ckelberg decomposition of the vector field,
$A_{\mu}={\tilde A}_{\mu}+\partial_{\mu}\lambda$ where $\partial_{\mu}{\tilde A}^{\mu}\equiv 0$,
in order to check that the longitudinal component
becomes a ghost \cite{Peloso1}. One could object \cite{GV} against a change of variables with time-derivatives. For if we make a
change of variables $x(t)\equiv{\dot y}(t)$ for a free non-relativistic particle we obtain an equation of motion
$\frac{d^4}{dt^4}y=\frac{d^3}{dt^3}x=0$ which clearly has more solutions than the initial theory admits. However,
in the latter example it is crucial that we have restricted the class of variations for our Lagrangian.
Not only $\delta x\equiv\delta\dot y$ but also $\delta y$ should be equal zero at the boundary of the
time interval. We only make such variations of the function $x(t)$ which integrate to zero over the given
time interval, $\int x(t)dt=0$. Under this restriction, any paths with ${\ddot x}=const$ bring the action
to its extremal value just as well as all the standard solutions with zero constant would do. Note that
nothing like this may happen in the St{\" u}ckelberg trick where the class of variations remains intact.
Therefore, this type of critique is unsubstantiated; and indeed, a careful analysis \cite{Peloso2, Lyth} shows that
the longitudianl mode {\it is} a ghost for short wavelengths.

The relevance of the ghost problem for models with stable explicit potential and non-minimal coupling is arguable
because at small scales we are not really to trust the approximation in which the scalar curvature
is just a fixed (slightly time-dependent) quantity and acts as a contribution to the effective
mass-squared. However, this observation does not give a recipe for how to treat the quantum production
of the vector field fluctuations at sub-horizon scales. Recall that the ghost appears just right below
the horizon length-scale, at momentum-squared $\approx 2H^2$. And the fluctuations of metric should produce
regions with both signs of the scalar curvature. It was claimed in \cite{Lyth,Lyth3} that the theory is under
control, partly because the total energy in the cosmological background with small occupation numbers of quantum
fluctuation modes is positive. However, the ghosts are extremely dangerous because they are produced
with divergent cross-section in Lorentz-invariant theories, and it is unclear whether a deep sub-horizon
UV-cutoff could be helpful in achieving a viable cosmological evolution. Nevertheless, for super-horizon
evolution the $\delta N$ formalism was properly generalized and applied to vector inflationary models \cite{Lyth}.
And ignoring the short-wavelength problems, even non-Gaussianity from vector field perturbations
can be calculated \cite{Lyth2}.

The full quantum problem of longitudinal modes still requires a thorough investigation. It is remarkable
though that the classical evolution of vector fields is smooth, at least in the test field approximation
\cite{Peloso3, me}. Unfortunately, once the gravitational backreaction is taken into account, numerical integration
of the full set of equaitons of motion for linear perturbations leads to divergencies \cite{Peloso3}.
This effect is a bit counter-intuitive, see for example discussion in \cite{me}, and a good analytical
understanding is needed. However, taken as it is, it
throws a big shadow on any attempts to refer to the short-scale dynamics of the scalar curvature as a means
of resolving the ghost problem because it is presumably the metric fluctuations which have destabilized
the longitudinal vector field dynamics in \cite{Peloso3}. One can argue that we just have to understand
non-linear dynamics of the theory \cite{Lyth3}, and that can well be the case but then for now we have no tractable
model even at the level of classical dynamics. And even if we resolve this issue, we may still want to
get the normal sign of the mass term after inflation which requires a transition through a highly
singular point \cite{Peloso3, me} of $m^2_{eff}=0$.

Let us now turn to the second model of \cite{Ford}. It makes use of a very flat potential with the
right sign of $V^{\prime}$ where the derivative is taken with respect to $A_{\mu}A^{\mu}=A_0^2-
a^{-2}{\overrightarrow A}^2$. In the $B$-variables, $B_0=A_0$, ${\overrightarrow B}=a^{-1}{\overrightarrow A}$,
we have a function of $B^2\equiv B_0^2-{\overrightarrow B}^2$.
The argument rolls very fast but the function changes only slightly.

One
possible example of such potential is $V=-\frac{B^2}{|B^2|^{1-\epsilon}}$
for small values of $\epsilon$. This is non-analytic but we don't worry about the neighbourhood of
zero and work only with negative values of $B^2$. In this case $V^{\prime}<0$ and therefore there is no ghost.
Let us however choose a background solution as follows: $B_{\mu}=\delta^{1}_{\mu}{\mathcal B}(t)$ and expand the
potential in terms of the longitudinal vector field fluctuation $\delta B_{\mu}=\partial_{\mu}\lambda$
up to the second order terms. We have in Lagrangian:
$$-V=\frac{-{\mathcal B}^2-2{\mathcal B}\partial_1 \lambda+(\partial_{\mu}\lambda)(\partial^{\mu}\lambda)}
{\left({\mathcal B}^2+2{\mathcal B}\partial_1 \lambda-(\partial_{\mu}\lambda)(\partial^{\mu}\lambda)\right)^{1-\epsilon}}=
{\mathcal B}^{2\epsilon}\left(-1-2\epsilon\frac{\partial_1\lambda}{\mathcal B}+\epsilon
\frac{(\partial_{\mu}\lambda)(\partial^{\mu}\lambda)}{{\mathcal B}^2}+
(2\epsilon-2\epsilon^2)\frac{(\partial_1\lambda)^2}{{\mathcal B}^2}\right).$$
Although the coefficients are strongly time-dependent, we see that for small values of $\epsilon$
the longitudinal mode exhibits a gradient instability. Only the models with $\epsilon\geqslant\frac12$
are free of this problem but they have potential energy which decays faster than $\frac{1}{a}$.

A working example of a very flat potential is $V=C \left(e^{\kappa\sqrt{|B^2|}}-1\right)$.
It has $V^{\prime}<0$ for $B^2<0$, and in principle can be smoothed around $B^2=0$ and
continued to $B^2>0$ in a healthy way by changing the sign in front of the square root. In this
model the gradient instability does not occur because after Taylor expanding
$$-V=C\left(1-e^{\kappa {\mathcal B}}\right)+Ce^{\kappa {\mathcal B}}\left(-\kappa\partial_1\lambda+
\frac{\kappa}{2{\mathcal B}}(\partial_{\mu}\lambda)(\partial^{\mu}\lambda)-
\left(\frac{\kappa^2}{2}-\frac{\kappa}{2{\mathcal B}}\right)(\partial_1\lambda)^2\right)+{\mathcal O}(\lambda^3)$$
we always have the correct sign in front of $(\partial_1\lambda)^2$.

In general we obtain $-\delta V=-V^{\prime}\left(2{\mathcal B}\partial_1\lambda+(\partial_{\mu}\lambda)(\partial^{\mu}\lambda)\right)-
2V^{\prime\prime}{\mathcal B}^2(\partial_1\lambda)^2+{\mathcal O}(\lambda^3)$. Therefore, we see that for stability we need
$V^{\prime}<0$ and an additional inequality
\begin{equation}
\label{stable}
V^{\prime\prime}\geqslant\frac{V^{\prime}}{2{\mathcal B}^2}
\end{equation}
to be satisfied. In particular, one can check that the
potential considered by Ford \cite{Ford},
$V=C\left(1-e^{\kappa B^2}\right)$, is
stable with respect to the gradient instability only when ${\mathcal B}^2\geqslant\frac{1}{2\kappa^2}$.

We see that the best we could do without introducing the ghosts is to choose a potential insensitive to the fast roll
of the $B$-field, but not to preserve $B_i$ from decaying. However, we should point out
that in constructing the most general vector field models, the problem is not to make
$A_i$ growing or $B_i$ stable. It can be achieved by a simple field redefinition. The real problem
is to have a slowly changing potential energy for a more natural model than that with such
an extremely flat potential. The latter task will be discussed later, and now we would like
to desrcibe a simple vector field redefinition which gives the growth of the field variable
but makes the Hamiltonian analysis
significantly harder without changing the physical content of the theory. The results are not
surprising of course but we find this calculation instructive and useful.

\section{A change of variables}

We start with a very simple action
\begin{equation}
\label{standard}
S=\int d^4 x \left(-\frac14 F_{\mu\nu}F^{\mu\nu}+\frac12 m^2 A_{\mu} A^{\mu}\right)
\end{equation}
for a (Proca) vector field in Minkowski space with the field strength $F_{\mu\nu}\equiv\partial_{\mu}A_{\nu}-\partial_{\nu}A_{\mu}$
and the metric signature ${+,-,-,-}$. Let us perform a change of variables such that
\begin{equation}
\label{change}
A_{\mu}=f({\mathfrak A}^2)\cdot {\mathfrak A}_{\mu}
\end{equation}
where $f$ is a function of the scalar argument ${\mathfrak A}^2\equiv {\mathfrak A}_{\mu}{\mathfrak A}^{\mu}$.
This is just a non-derivative one-to-one change of variables
whenever $f \neq 0$ and $f+2f^{\prime}{\mathfrak A}^2\neq 0$, and we expect of course that the physical Hamiltonian
in terms of the new variables can be obtained just by the change of variables (\ref{change}) in the physical
Hamiltonian for the action (\ref{standard}).
However, a straightforward Hamiltonian analysis
turns out to be remarkably complicated for this very simple theory in unusual variables.

We outline the main steps of canonical analysis for the aciton
\begin{equation}
\label{changed}
S=\int d^4 x \sqrt{-g} \left(-\frac14
\left(\partial_{\mu}(f({\mathfrak A}^2){\mathfrak A}_{\nu})-\partial_{\nu}(f({\mathfrak A}^2){\mathfrak A}_{\mu})\right)
\left(\partial^{\mu}(f({\mathfrak A}^2){\mathfrak A}^{\nu})-\partial^{\nu}(f({\mathfrak A}^2){\mathfrak A}^{\mu})\right)
+\frac{m^2}{2} f^2({\mathfrak A}^2) {\mathfrak A}_{\mu} {\mathfrak A}^{\mu}\right)
\end{equation}
of the {\it standard} massive vector field in the new variables. We notice first that
there is a simple relation for the velocities:
$${\dot A}_i=f{\dot {\mathfrak A}}_i+2f^{\prime}{\mathfrak A}_i({\mathfrak A}_0{\dot {\mathfrak A}}_0-{\mathfrak A}_k{\dot {\mathfrak A}}_k)$$
where, for the sake of brevity, we have omitted the argument $({\mathfrak A}^2)$ of the
function $f$ and its derivative. It gives for canonical momenta
${\mathfrak P}^{\mu}\equiv\frac{\partial {\mathcal L}}{\partial {\dot {\mathfrak A}}_{\mu}}=
\frac{\partial {\mathcal L}}{\partial {\dot A}_{\nu}}\cdot \frac{\partial {\dot A}_{\nu}}{\partial {\dot {\mathfrak A}}_{\mu}}=
P^{\nu}\frac{\partial {\dot A}_{\nu}}{\partial {\dot {\mathfrak A}}_{\mu}}$ the following expressions:
\begin{eqnarray}
\label{momenta}
{\mathfrak P}^{0}=2f^{\prime}{\mathfrak A}_0 {\mathfrak A}_k\left({\dot A}_k-\partial_k A_0\right);&
\qquad{\mathfrak P}^{i}=\left(f\delta_{ik}-2f^{\prime}{\mathfrak A}_i {\mathfrak A}_k\right)\left({\dot A}_k-\partial_k A_0\right).
\end{eqnarray}

We easily observe that ${\mathfrak A}_i {\mathfrak P}^i=\left(f-2f^{\prime}{\mathfrak A}_k^2\right)
{\mathfrak A}_i\left({\dot A}_i-\partial_i A_0\right)$, and deduce the primary constraint:
\begin{equation}
\label{primary}
{\mathfrak P}^{0}-\frac{2f^{\prime}{\mathfrak A}_0}{f-2f^{\prime}{\mathfrak A}_k^2}{\mathfrak A}_i {\mathfrak P}^i=0.
\end{equation}
Of course, in the old variables it can be expressed just as $P^0=0$. To check this fact we have to
invert the Jacobian matrix $\frac{\partial {\dot A}_{\mu}}{\partial {\dot {\mathfrak A}}_{\nu}}=
f\delta_{\mu}^{\nu}+2f^{\prime}{\mathfrak A}_{\mu}{\mathfrak A}^{\nu}$. The answer is
\begin{equation}
\label{invert}
\frac{\partial {\dot {\mathfrak A}}_{\nu}}{\partial {\dot A}_{\mu}}=\frac{1}{f}
\left(\delta_{\nu}^{\mu}-\frac{2f^{\prime}{\mathfrak A}_{\nu}{\mathfrak A}^{\mu}}{f+2f^{\prime}{\mathfrak A}^2}\right).
\end{equation}
And we readily see that
$$P^0=\frac{\partial {\dot {\mathfrak A}}_{\nu}}{\partial {\dot A}_{0}}{\mathfrak P}^{\nu}=
\frac{1}{f\left(f+2f^{\prime}{\mathfrak A}^2\right)}
\left({\mathfrak P}^{0}\left(f-2f^{\prime}{\mathfrak A}_k^2\right)-2f^{\prime}{\mathfrak A}_0{\mathfrak A}_i {\mathfrak P}^i\right)$$
is proportional to the constraint (\ref{primary}).

Let's use the primary constraint (\ref{primary}) in the Hamiltonian $H={\mathfrak P}^{\mu}
{\dot {\mathfrak A}}_{\mu}-{\mathcal L}$ to obtain
$$H={\mathfrak P}^i\left({\dot {\mathfrak A}}_{i}+
\frac{2f^{\prime}{\mathfrak A}_0{\dot {\mathfrak A}}_0}{f-2f^{\prime}{\mathfrak A}_k^2}{\mathfrak A}_i\right)
-\frac12\left({\dot A}_i-\partial_i A_0\right)^2+
\frac14\left(\partial_i\left(f{\mathfrak A}_k\right)-\partial_k\left(f{\mathfrak A}_i\right)\right)^2
-\frac12 m^2 f^2{\mathfrak A}^2.$$
We need to get rid off velocities. Using the definition of momenta we have
$$\left({\mathfrak P}^i\right)^2=f^2\left({\dot A}_i-\partial_i A_0\right)^2+
4\left({f^{\prime}}^2{\mathfrak A}_k^2-ff^{\prime}\right)
\left({\mathfrak A}_i\left({\dot A}_i-\partial_i A_0\right)\right)^2$$
and then
$$\frac12\left({\dot A}_i-\partial_i A_0\right)^2=\frac{1}{2f^2}\left({\mathfrak P}^i\right)^2-
2\frac{{f^{\prime}}^2{\mathfrak A}_k^2-ff^{\prime}}{f^2\left(f-2f^{\prime}{\mathfrak A}_m^2\right)^2}
\left({\mathfrak A}_i {\mathfrak P}^i\right)^2=\frac{1}{2f^2}\left(\left(\delta_{ik}+
\frac{2f^{\prime}{\mathfrak A}_i{\mathfrak A}_k}{f-2f^{\prime}{\mathfrak A}_m^2}\right){\mathfrak P}^k\right)^2.$$
With some simple algebra we also transform the first term in the Hamiltonian
\begin{multline*}
{\mathfrak P}^i\left({\dot {\mathfrak A}}_{i}+
\frac{2f^{\prime}{\mathfrak A}_0{\dot {\mathfrak A}}_0}{f-2f^{\prime}{\mathfrak A}_k^2}{\mathfrak A}_i\right)=
\frac{{\mathfrak P}^i}{f}\left({\dot A}_i+
\frac{2f^{\prime}{\mathfrak A}_i{\mathfrak A}_k {\dot A}_k}{f-2f^{\prime}{\mathfrak A}_m^2}\right)=\\
=\frac{{\mathfrak P}^i}{f}\left(\left({\dot A}_i-\partial_i A_0\right)+
\frac{2f^{\prime}{\mathfrak A}_i{\mathfrak A}_k \left({\dot A}_k-\partial_k A_0\right)}{f-2f^{\prime}{\mathfrak A}_m^2}\right)+
\frac{{\mathfrak P}^i}{f}\left(\partial_i A_0+
\frac{2f^{\prime}{\mathfrak A}_i{\mathfrak A}_k \partial_k A_0}{f-2f^{\prime}{\mathfrak A}_m^2}\right),
\end{multline*}
of which the first part yields the expression we already know (we use (\ref{momenta})):
\begin{multline*}
\frac{{\mathfrak P}^i}{f}\left(\left({\dot A}_i-\partial_i A_0\right)+
\frac{2f^{\prime}{\mathfrak A}_i{\mathfrak A}_k \left({\dot A}_k-\partial_k A_0\right)}{f-2f^{\prime}{\mathfrak A}_m^2}\right)=\\
=\left(\left({\dot A}_i-\partial_i A_0\right)-2\frac{f^{\prime}}{f}{\mathfrak A}_i {\mathfrak A}_k\left({\dot A}_k-\partial_k A_0\right)\right)
\cdot\left(\left({\dot A}_i-\partial_i A_0\right)+
\frac{2f^{\prime}{\mathfrak A}_i{\mathfrak A}_k \left({\dot A}_k-\partial_k A_0\right)}{f-2f^{\prime}{\mathfrak A}_m^2}\right)=\\
=\left({\dot A}_i-\partial_i A_0\right)^2.
\end{multline*}
This is twice the contribution from the Lagrangian density and has the opposite sign, precisely as it should be
for the quadratic in momenta part of the Hamiltonian given that the Lagrangian was quadratic
in velocities. Finally, we combine everything together and get the Hamiltonian density:
\begin{multline}
\label{Hamiltonian}
H=\frac{1}{2f^2}\left(\left(\delta_{ik}+
\frac{2f^{\prime}{\mathfrak A}_i{\mathfrak A}_k}{f-2f^{\prime}{\mathfrak A}_m^2}\right){\mathfrak P}^k\right)^2+
\frac{{\mathfrak P}^i}{f}\left(\partial_i \left( f{\mathfrak A}_0\right)+
\frac{2f^{\prime}{\mathfrak A}_i{\mathfrak A}_k \partial_k \left( f{\mathfrak A}_0\right)}{f-2f^{\prime}{\mathfrak A}_m^2}\right)+\\
+\frac14\left(\partial_i\left(f{\mathfrak A}_k\right)-\partial_k\left(f{\mathfrak A}_i\right)\right)^2
+\frac12 m^2 f^2{\mathfrak A}_i^2-\frac12 m^2 f^2{\mathfrak A}_0^2.
\end{multline}

This is actually the usual Hamiltonian $H=\frac12 \left(P^i\right)^2-A_0\partial_i P^i+\frac14 F_{ik}^2+\frac12 m^2
\left(A_i^2-A_0^2\right)$ in the new variables. Indeed, the first terms in the Hamiltonians do
coincide because
$$P^i=\frac{\partial {\dot {\mathfrak A}}_{\nu}}{\partial {\dot A}_{i}}{\mathfrak P}^{\nu}=
\frac{1}{f}\left(\delta_{ik}+
\frac{2f^{\prime}{\mathfrak A}_i{\mathfrak A}_k}{f-2f^{\prime}{\mathfrak A}_m^2}\right){\mathfrak P}^k$$
where we have used equation (\ref{invert}) and the primary constraint (\ref{primary}). And now
we also see that after integration by parts the second term in (\ref{Hamiltonian}) becomes
$-f{\mathfrak A}_0\partial_i P^i=-A_0\partial_i P^i$.

The next step is to find the secondary constraint. We know that after canonical transformation
the Poisson brackets should not have changed, and therefore the secondary constraint must
acquire a form equivalent to $\partial_i P^i+m^2 A_0=0$. But it's not an easy task to
establish this result with a straightforward computation of the Poisson bracket of the
Hamiltonian (\ref{Hamiltonian}) with the primary constraint (\ref{primary}). We can however
explicitly check that for $P^0\equiv
\frac{1}{f\left(f+2f^{\prime}{\mathfrak A}^2\right)}
\left({\mathfrak P}^{0}\left(f-2f^{\prime}{\mathfrak A}_k^2\right)-2f^{\prime}{\mathfrak A}_0{\mathfrak A}_i {\mathfrak P}^i\right)$
and $A_{\mu}\equiv f({\mathfrak A}^2)\cdot {\mathfrak A}_{\mu}$ the Poisson brackets are
$\{P^0,A_{\mu}\}=\frac{\partial P^0}{\partial{\mathfrak P}^{\alpha}}
\frac{\partial A_{\alpha}}{\partial{\mathfrak A}_{\mu}}=\delta^0_{\mu}$. To calculate the
$\{H,P^0\}$ quantity we need also the $\{P^0,P^i\}$ bracket which can be computed using the obvious
relation $P^{\mu}=\frac{\partial {\dot {\mathfrak A}}_{\nu}}{\partial {\dot A}_{\mu}}{\mathfrak P}^{\nu}$ together
with the formula (\ref{invert}). It's quite a bulky endeavor which can be simplified by observing
that the right hand side of (\ref{invert}) contains two scalar functions of ${\mathfrak A}^2$: $h\equiv\frac{1}{f}$
and $g\equiv\frac{2f^{\prime}}{f^2+2ff^{\prime}{\mathfrak A}^2}$. The Poisson bracket appears to be
$\{P^0,P^i\}=\left(2hh^{\prime}+hg-2g{\mathfrak A}^2h^{\prime}\right)
\left({\mathfrak A}_0{\mathfrak P}^i-{\mathfrak A}_i{\mathfrak P}^0\right)=0.$
And hence we have $\{H,P^0\}=\partial_i P^i+m^2 A_0=0$ which gives the secondary
constraint
\begin{equation}
\label{secondary}
\partial_i\left(\left(\delta_{ik}+
\frac{2f^{\prime}{\mathfrak A}_i{\mathfrak A}_k}{f-2f^{\prime}{\mathfrak A}_m^2}\right)
\frac{{\mathfrak P}^k}{f}\right)+m^2f{\mathfrak A}_0=0.
\end{equation}

We have a pair of second class constraints, and the canonical analysis stops. It is not however possible
to exclude the unphysical variable ${\mathfrak A}_0$ from the Hamiltonian explicitly due to a very complicated
form of equation (\ref{secondary}). Nevertheless, it's obvious that the Hamiltonian density is positive
definite and equals to $${\mathcal H}=\frac12 \left(P^i\right)^2+\frac14 F_{ik}^2+\frac12 m^2\left(A_i^2+A_0^2\right)=
\frac12 \left(P^i\right)^2+\frac{1}{2m^2}\left(\partial_i P^i\right)^2+\frac14 F_{ik}^2+\frac12 m^2A_i^2.$$
Note also that the apparent singularity at $f=2f^{\prime}{\mathfrak A}^2_i$
is unphysical and only reflects the fact that the momentum ${\mathfrak P}^0$ can not be always assumed unphysical.
Near this locus we should have rather excluded ${\mathfrak A}_i {\mathfrak P}^i$ than ${\mathfrak P}^0$.

Of course, we could do the same analysis with a general potential. And, clearly, we could search
for such a potential which will acquire the form of a mass term in the new variables. In this
case, due to the growth of the new vector field variables, the potential energy may become slowly rolling.
However, it will require a tachyonic potential in initial model. And now, separating the longitudinal modes,
we will make a St{\" u}ckelberg decomposition for the $f({\mathfrak A}^2){\mathfrak A}_{\nu}$ field which
would of course reveal the ghosty kinetic energy again. In other words, what we have seen in this Section
is {\it not} a modification of the vector field theory. And we have to test something different.

\section{Lorentz-invariant modifications}

Probably, the first idea after discussion in the previous Seciton is to take a combination of
two kinetic terms,
$$\mathcal L=-\frac14\left(cF_{\mu\nu}F^{\mu\nu}+{\tilde F}_{\mu\nu}{\tilde F}^{\mu\nu}\right)-V$$
where $F_{\mu\nu}\equiv\partial_{\mu}A_{\nu}-\partial_{\nu}A_{\mu}$ and
${\tilde F}_{\mu\nu}\equiv\partial_{\mu}(f(A^2)A_{\nu})-\partial_{\nu}(f(A^2)A_{\mu})$.
However, in this case we have the same momentum $p^0$ as in equation (\ref{momenta}) and
the spatial momenta $p^i$ are shifted by $cF_{0i}$, therefore we get the following relation
instead of the primary constraint (\ref{primary}):
$$2f^{\prime}A_0 A_i p^i-(f-2f^{\prime}A_k^2)p^{0}=2f^{\prime}cA_0 A_i({\dot A}_i-\partial_i A_0).$$
If $A_0\neq 0$ it allows to solve for $A_i{\dot A}_i$ in terms of fields and momenta. The temporal
momentum contains the time derivative of $f(A^2)A_i$ (see (\ref{momenta}) and recall that the Latin letters
are used to denote the variables which correspond to the Gothic fields of the previous Section), and therefore now we can
determine ${\dot A}_0$. Finally, we have enough equations from the definition of $p^i$'s to
solve for the two remaining velocities. Therefore, a model with different kinetic terms has an
ill-defined number of degrees of freedom which equals four almost everywhere. We would however prefer
a vector field with strictly three degrees of freedom.

The next possibility is to have a kinetic self-coupling, $f(A^2)F^2$. Let us take a mass-term potential
for simplicity
$${\mathcal L}=-\frac14 f(A^2)F_{\mu\nu}F^{\mu\nu}+\frac12 m^2A^2$$
and perform the Hamiltonian analysis. The canonical momenta are simply given by $p^i=fF_{0i}$ and $p^0=0$.
The Hamiltonian density $${\mathcal H}=\frac{(p^i)^2}{2f}-A_0\partial_i p^i+\frac14 F_{ik}F_{ik}-\frac12 m^2A^2$$
leads to the secondary constraint $-\partial_i p^i=m^2A_0+\frac{(p^i)^2f^{\prime}}{f^2}A_0
-\frac12 f^{\prime}A_0F_{ik}F_{ik}$ which can not be explicitly solved but allows to write the Hamiltonian
in the following form:
$${\mathcal H}=\frac{(p^i)^2}{2f}\left(1+2\frac{f^{\prime}}{f}A_0^2\right)+
\frac14 F_{ik}F_{ik}\left(1-2\frac{f^{\prime}}{f}A_0^2\right)+\frac12 m^2\left(A_0^2+A_i^2\right)$$
which is bounded neither from below nor from above for any non-constant function $f$. This is a
short-wavelength
problem, and therefore can be dangerous. (A little thought shows that this result
is generic also for non-linear non-trivial functions of two arguments, $A^2$ and $F^2$.)
And let us look at the equations of motion:
$$\bigtriangledown_{\mu}(fF^{\mu\nu})-\frac12 f^{\prime}F^2A^{\nu}+m^2A^{\nu}=0.$$
If we are searching for a slow-roll solution with negligible time-dependence of the function $f$, then
we effectively have a massive vector field of mass ${\tilde m}^2=\frac{m^2}{f}-\frac{f^{\prime}F^2}{2f}$
with $F^2\approx H^2B^2$. As $F^2<0$ and $f>0$, we must have $f^{\prime}<0$ and
$|f^{\prime}F^2|>2m^2$ for the mass to be negative (let alone being close to $-2H^2$).
Having obtained these inequalities, we check the quadratic term in the longitudinal mode action ${\mathcal L}=
\left(-\frac14 f^{\prime}F^2+\frac12 m^2\right)(\partial_{\mu}\lambda)(\partial^{\mu}\lambda)$,
and find a ghost again.

We could also try to modify the $F_{\mu\nu}F^{\mu\nu}$ structure of kinetic function. However, it is
easy to see that there are only two quadratic ghost-free possibilities: the standard one
and $(\partial_{\mu}A^{\mu})^2$ which propagates only one degree of freedom. Relaxing the
condition of being quadratic, we get one more possible structure: $G_{\mu\nu}=
A^{\alpha}\left(A_{\nu}\partial_{\mu}A_{\alpha}-A_{\mu}\partial_{\nu}A_{\alpha}\right)$
which is naturally produced in a particular
combination with $F_{\mu\nu}$ by the change of variables (\ref{change}).
Let us now take the most general kinetic part of Lagrangian
quadratic in $F$ and $G$:
$${\mathcal L}=-\frac14 \left(f(A^2)F_{\mu\nu}F^{\mu\nu}
+2g(A^2)F_{\mu\nu}G^{\mu\nu}+h(A^2)G_{\mu\nu}G^{\mu\nu}\right).$$
The canonical momenta are:
\begin{eqnarray}
\label{momenta2}
p^0=A_0 A_k\left(gF_{ok}+hG_{ok}\right);&
\qquad p^{i}=\left(fF_{oi}+gG_{oi}\right)-A_i A_k\left(gF_{ok}+hG_{ok}\right).
\end{eqnarray}
So that we easily find a simple relation
\begin{equation}
\label{relation}
p^i=fF_{oi}+gG_{oi}-\frac{A_i}{A_0}p^0
\end{equation}
and see that
if
\begin{equation}
\label{three}
\frac{f}{g}=\frac{g}{h}
\end{equation}
then it can be written entirely in terms of momenta
without velocities, and therefore in this case we find a primary constraint.
Otherwise we can combine the relation (\ref{relation}) with the definition
of the temporal momentum in (\ref{momenta2}) and determine both
$A_kF_{0k}$ and $A_kG_{0k}$ in terms of momenta if $A_0\neq 0$. It gives us $A_i{\dot A}_i$
and ${\dot A}_0$. And remaining two independent equations in (\ref{momenta2}) allow
us to find the two remaining velocities. Hence, equation (\ref{three}) is
the necessary and sufficient condition to have a vector field with three degrees
of freedom. However, in this case our Lagrangian is simply
${\mathcal L}=-\frac14 (fF+hG)^2$ and it can always be represented as a change of variables
(\ref{change}) in a Lagrangian of $-\frac14 f^2F^2$ type (probably, with constant $f$) which
we have studied above. Indeed, we want to convert $fF+hG$ into ${\tilde f}({\tilde h}F+2{\tilde h}^{\prime}G)$ for what
it suffices to put ${\tilde h}=\exp\int\frac{h}{2f}$ and ${\tilde f}=\frac{f}{\tilde h}$.

We also have a possibility of taking non-linear functions of simple kinetic terms,
for example ${\mathcal L}=-f(F^2)-V(A^2)$. An accelerating solution for
$f(F^2)=\frac{F^2}{4}-\frac{c}{F^2}$ with negative constant $c$ was constructed in \cite{Novello}.
However, this model has an ill-defined number of degrees of freedom at $f^{\prime}=0$, and
also it has a Hamiltonian unbounded from below. One can actually check
that $f^{\prime}>0$ and $V^{\prime}<0$ are necessary conditions for the Hamiltonian
to be bounded from below \cite{Esposito}. Unfortunately, those theories of this class which can
give an interesting dynamics are necessarily unstable. If a vector field is to play any significant role
in the cosmological expansion, then some terms in its Lagrangian should not be diluted.
Unless either $V$ or $f$ is an extremely flat function (in the latter case the transverse vector
fluctuations would be strongly coupled),
it means that either $\frac{A_i}{a}$ or $\frac{F_{0i}}{a}$ have to roll slowly. In \cite{Esposito} it
was shown that neither of this options is available whenever $f^{\prime}>0$ and $V^{\prime}<0$.
Introduction of an additional $F{\tilde F}$ argument
with dual field strength tensor $\tilde F$ to the function $f$ does not change the cosmological
dynamics \cite{Esposito}, as the dual tensor has only spatial non-vanishing components. An extra
$\bigtriangledown_{\mu}A^{\mu}$ argument would generically lead to an extra (fourth) degree
of freedom.

In principle one could consider a non-linear function of several arguments $f(F^2,FG,G^2,A^2)$. However, in order for
the momenta to satisfy a linear constraint equation we would need a condition analogous to
(\ref{three}) with $\frac{\partial f}{\partial F^2}$, $\frac{\partial f}{\partial FG}$ and
$\frac{\partial f}{\partial G^2}$ instead of $f$, $2g$ and $h$ respectively, and the ratios
being independent of kinetic arguments. It follows then that a linear in $F^2$, $FG$, and $G^2$
increment of the function $f$ is always propotional to $\left(c_1(A^2)F+c_2(A^2)G\right)^2$,
and again we have got no new options.

To summarize, no new viable models of vector inflation have been found in this Section. We can not completely exclude
a possibility that there may be some highly non-linear models of very clever
design in the class of Lorentz-invariant vector field theories with three propagating degrees of freedom
which would be able to produce a vector inflationary regime. However, it is clear that
generically this is not possible, except for vector fields with extremely flat potentials
as was proposed in the paper \cite{Ford}. (We should stress again that, throughout the paper, only
Abelian vector fields are being discussed. In Refs. \cite{Sheikh1,Sheikh2} an inflationary model is
constructed with a special $F^4$-correction to the Yang-Mills action. The model is explicitly
gauge invariant, and therefore may be free of some problems. However, a full Hamiltonian analysis
was not yet performed. And negative values of sound speed squared for some fluctuation modes
are reported \cite{Sheikh2}. It signals a gradient instability of the background solution.)

\section{A few remarks on Lorentz-breaking models}

It is of course the restrictions of Lorentz-invariance which prevent us from constructing a suitable
model. And it is also the Lorentz-invariance which makes the ghosts so dangerous. The standard argument
is as follows. Consider a graviton-mediated creation of a pair of normal particles and a pair of
ghosts from nothing. Let's fix a reference frame. One possible kinematics of this decay is the one with compensated
spatial momenta inside every pair (a pair of identical particles with precisely opposite momenta,
and an analogous pair of ghosts), such that the negative energy of ghosts compensates
the positive energy of particles. However, there are lots of other possibilities too.
The vacuum decay can proceed with a kinematics which looks in the chosen frame precisely as
the pair-momentum-compensated kinematics would have looked like in some other, Lorentz-boosted frame.
And therefore, in calculating the amplitude we would have to integrate over
all possible momenta and over all possible Lorentz frames. It gives a divergent result
regardless of how small the coupling is. The kinematics of the vector ghost is somewhat different.
And in the ultraviolet limit even a tiny portion of transverse excitation can
compensate the negative longitudinal energy \cite{me}. But nevertheless, at the very least,
the infinite Lorentz-group volume is unavoidable.
In Lorentz-violating models the rate of the ghost production could in principle be controlled,
but it remains to be understood whether an ultraviolet cutoff deep under the Hubble length-scale
could be helpful for cosmology.

A natural way to proceed with the Lorentz-breaking scenarios is to invoke
Lorentz-breaking vector potentials which could come not only from theories
with fundamentally preferred frames but also from couplings to an aether
field or some non-trivial background; a small breaking can even occur
due to $R^{\mu\nu}A_{\mu}A_{\nu}$ coupling. A simple example is
$V(A^2)=-(m_1^2A_0^2-m_2^2A_i^2)$ with different masses for temporal
and longitudinal components.
In \cite{me} it was shown that at the classical level
a major analytic problem of tachyonic vector field comes from the temporal component
of equations of motion which is normally used to determine unphysical $A_0$ variable,
$$\left(-\bigtriangleup+m^2\right)A_0+\partial_i{\dot A}_i=0.$$
With tachyonic mass, the spectrum of the operator in front of $A_0$ contains zero.
We can in principle overcome this trouble if we take $m_1^2>0$ and $m_2^2<0$.
(The Hamiltonian is then unbounded from below but only due to tachyonic effect which
can be cured by a non-linear potential for spatial components.)
This is not easy to
do with coupling to the Ricci tensor because in quasi-de-Sitter regime it is
almost proportional to the metric. One can check that in order to have $m_2^2$ of
order $-2H^2$ and $m_1^2$ positive, we will need to use couplings to Ricci scalar
and Ricci tensor of order ${\mathcal O}\left(\frac{1}{\epsilon}\right)$ where
$\epsilon$ is the slow-roll parameter, $\epsilon\equiv-\frac{\dot H}{H^2}$. This difference of masses may be
better achieved with a coupling to an aether field. However, it would solve only the ghost problem,
but the gradient istabilities would persist. It is evident from the St{\" u}ckelberg analysis,
and can also be seen directly from equations of motion (in Minkowski space):
$${\ddot A}_i+m_2^2A_i-\bigtriangleup A_i+\left(1-\frac{m_2^2}{m_1^2}\right)\partial_i\partial_k A_k=0.$$
The gradient instability is dangerous because the fluctuation modes grow with an unbounded rate in
the ultraviolet, and even the condition of bounded energy of the initial fluctuation does
not help to control this process. If however we are dealing with a theory which contains higher spatial
derivatives, like in Ho{\v r}ava gravity \cite{Horava} for example, then the rate of fluctuation growth
may become bounded if the higher derivative terms go with proper signs. This issue deserves a further
investigation.

There is also a temptation to modify the kinetic term. It is a fairly simple task if
there are no restrictions on the
choice to be made. A model with no temporal components
${\mathcal L}=\frac12 \left(\partial_{\mu}A_i\right)(\partial^{\mu}A_i)-\frac12 m^2A_i^2$ would be
just perfect, but hardly of any physical interest. Note however that there are
less radical choices too. For example, a theory with ${\mathcal L}=-f^2({\overrightarrow A}^2)F^2-V(A^2)$
obviously enjoys a Hamiltonian bounded from below if the potential is stable. Note also
that a peculiar modification of kinetic term can be deduced as a dimensional reduction
of 5-dimensional gravity with 4-dimensional Lovelock (Gau{\ss}-Bonnet) invariant, see \cite{Esposito}.
It is an open task to explore the cosmological consequences of this model. But any further
discussion of Lorentz-violating vector field theories is beyond the scope of the present paper.

\section{Conclusions}

Vector inflation was invented as an alternative to the scalar models and received a considerable
interest among cosmologists partly because it could give a nice account of
possible asymmetries in the CMB \cite{erik}. At the background level and at the super-horizon
scales the model is tractable, and allowed for $\delta N$ calculations \cite{Lyth, Lyth2}.
However, the full linear perturbation equations are cumbersome \cite{GV}, and the generic models
are known to be badly unstable \cite{Peloso1, me}. In this paper we have found out that in the
class of Lorentz-invariant vector field models with three degrees of freedom, the ghost problem
can not be resolved unless the potential is taken to be an extremely flat function
as was proposed in probably the very first paper on the subject \cite{Ford}. But there are many
Lorentz-breaking possibilities which are still waiting for a careful investigation.

\end{document}